\documentclass[reqno,10pt]{article}

\usepackage[english]{babel}
\usepackage[utf8]{inputenc}
\usepackage{fec}
\usepackage[a4paper, total={6.5in, 8.6in}]{geometry}
\usepackage[ruled, algo2e]{algorithm2e} 

\usepackage{subcaption}
\usepackage{tikz}

\usepackage[normalem]{ulem}
\usepackage{mathtools}
\usepackage{amsmath}
\usepackage{graphicx}
\usepackage{booktabs}
\usepackage{bbm}
\usepackage{todonotes}
\usepackage{xcolor}

\newtheorem*{example*}{Example: The Ring}

\newtheorem*{conjecture*}{Fixed Angle Conjecture}

\newtheorem*{task*}{Task}

\usepackage{pgfplots}

\usepackage[scr=rsfs]{mathalpha}

\usepackage{authblk}

\newcommand{\bra}[1]{\left\langle{#1}\right|}
\newcommand{\ket}[1]{\left|{#1}\right\rangle}

 \usetikzlibrary{angles, quotes}

\newcommand{\Bibkeyhack}[3]{}
\newcommand{\gambf}{\ensuremath{\boldsymbol{\gamma}}}
\newcommand{\betbf}{\ensuremath{\boldsymbol{\beta}}}

\newcommand\Item[1][]{%
  \ifx\relax#1\relax  \item \else \item[#1] \fi
  \abovedisplayskip=0pt\abovedisplayshortskip=0pt~\vspace*{-\baselineskip}}

\title{Strategies for running the QAOA at hundreds of qubits}

\author[1]{Brandon Augustino}  
\author[2]{Madelyn Cain}
\author[3]{Edward Farhi}
\author[1]{Swati Gupta}
\author[ ]{Sam Gutmann}
\author[4,5]{Daniel Ranard}
\author[4]{Eugene Tang}
\author[2]{Katherine Van Kirk}

\affil[1]{\textit{\footnotesize{Sloan School of Management, Massachusetts Institute of Technology, Cambridge, MA 02139}}}
\affil[2]{\textit{\footnotesize{Department of Physics, Harvard University, Cambridge, MA 02138}}}
\affil[3]{\textit{\footnotesize{Google Quantum AI, Venice, CA 90291}}}
\affil[4]{\textit{\footnotesize{Center for Theoretical Physics, Massachusetts Institute of Technology, Cambridge, MA 02139}}}
\affil[5]{\textit{\footnotesize{Walter Burke Institute for Theoretical Physics, California Institute of Technology, Pasadena, CA 91125}}}
 
\date{\today}

\begin{document}
\maketitle

\begin{abstract}

We explore strategies aimed at reducing the amount of computation, both quantum and classical, required to run the Quantum Approximate Optimization Algorithm (QAOA). 
We focus on the example of MaxCut on random $3$-regular graphs with up to a few hundred vertices. 

First, following Wurtz \emph{et al.}~\cite{wurtz2021fixed}, we consider the standard QAOA with instance-independent ``tree'' parameters chosen in advance. At depth $p$ these tree parameters are chosen to optimize the MaxCut expectation for graphs with girth at least $2p+1$. We provide extensive numerical evidence supporting the performance guarantee   for tree parameters conjectured in~\cite{wurtz2021maxcut} and see that the approximation ratios obtained with tree parameters are typically well beyond the conjectured lower bounds. Further, by using tree parameters as the initial point of a variational optimization, we typically see only slight improvements and get results comparable to performing a full optimization. This suggests that in practice, the QAOA can achieve near-optimal performance without the need for parameter optimization.

Next, we modify the warm-start QAOA of Tate \textit{et al.}~\cite{tate2023customMixers}. The starting state for the QAOA is now an optimized product state associated with a solution of the Goemans-Williamson (GW) algorithm. Surprisingly,  the tree parameters above continue to perform well for the warm-start QAOA. We find that for random 3-regular graphs with hundreds of vertices, the expected cut obtained by the warm-start QAOA at depth $p \gtrsim 3$ is at least comparable to that of the standard GW algorithm. 

Our numerics on random instances do not provide general performance guarantees but do provide substantial evidence that there exists a regime of instance sizes in which the QAOA finds good solutions at low depth without the need for parameter optimization. 
For each instance studied, we  classically compute the expected size of the QAOA distribution of cuts; producing the actual cuts requires running on a quantum computer. Exploring these strategies on fault-tolerant hardware, at or beyond the system sizes and depths we have considered, might provide new insights into the performance of quantum algorithms in regimes beyond what can be studied classically.

\end{abstract}

%\tableofcontents

%\newpage

\section{Introduction}
The Quantum Approximate Optimization Algorithm (QAOA)~\cite{farhi2014qaoa} can be used to find good approximate solutions to the problem of maximizing a cost function defined on bit strings. 
The standard QAOA consists of $p$ layers of alternating applications of cost-function-dependent unitaries and single-qubit rotations. Previous studies have often focused on either asymptotically large sizes or alternatively the near-term noisy intermediate-scale quantum (NISQ) era. 
This paper considers strategies for running the QAOA on graph problem sizes of, say, $500$ vertices, and at values of $p$ that may be relevant for running the QAOA on fault-tolerant hardware that may be available in the future. 
Although classical simulation at hundreds of qubits can be used to give the expected performance of the QAOA, one still needs to run it on hardware to obtain an associated string. 
While there is no known example where the QAOA beats all existing classical algorithms and classical heuristics at these system sizes perform  well, there is still much to be learned at depths and system sizes that may be achievable in the near future.

At any given depth $p$ the QAOA has $2p$ parameters which need to be established in order to run the algorithm. 
Good parameters can be searched for variationally, but this may require many applications of the quantum computer.  
For small values of $p$ there are cases where the parameters can be determined in advance by classical means,  thus avoiding this expensive quantum computer search. 
For example, consider the \textit{maximum cut} (MaxCut) problem on bounded degree graphs at low $p$. 
The classical search for parameters can be done anew for each graph and can be made efficient by considering small subgraphs corresponding to the neighborhoods of each edge.
While this search is a good strategy at low depth, the computational cost grows exponentially in $p$.  
For large graphs chosen from a distribution, it has been shown that for fixed QAOA parameters the expected value of the cost function varies little over different graphs \cite{brandao2018fixed}. 
This implies that optimal parameters for one graph will work for others from the same distribution.  
In the case of the Sherrington-Kirkpatrick model, where the ensemble is over different couplings on the complete graph, this landscape independence has been proven as well~\cite{farhi2022qaoaSK}.

We explore two strategies previously put forward by other groups for running the QAOA. 
The first strategy obviates the need for parameter search, while the second aims to improve the performance of the QAOA. 
By combining the two approaches, we obtain a modified strategy which appears to give good performance at the scale of hundreds of qubits. Our results are numerical and we do not offer proofs of performance.
%There are few quantum approaches to combinatorial optimization, and the QAOA is an easy to implement all-purpose quantum algorithm, which may work well in practice. 
%On the classical side, it can be compared to simulated annealing, which works well in practice even when performance guarantees cannot be established.  
We focus on MaxCut on random 3-regular graphs.
All of the results in this paper concern the QAOA acting on this ensemble, and we can only speculate about how these results apply in other situations.  

%We explore two discoveries in the literature. 
The first strategy we explore is based on the discovery  that, for  3-regular graphs at depth $p$, there are universal parameters that work well for any graph and can be found in advance \cite{wybo2024missing, wurtz2021fixed,wurtz2021maxcut}. 
These parameters are the optimal fixed $p$ parameters as the graph size goes to infinity when all edge neighborhoods look locally tree-like. Surprisingly, these parameters give high approximation ratios even for finite-size graphs which are not locally tree-like. 
This observation is related to a conjecture in \cite{wurtz2021maxcut} and all of our simulations support this conjecture. This leads to a strategy for running the QAOA at hundreds of qubits on 3-regular graphs: choose the desired depth $p$, determine the tree parameters using a classical computer, then use these predetermined parameters to run the QAOA on a quantum computer.  
These parameters are given in \cite{wurtz2021fixed} up to $p=11$.

The other strategy that we investigate is based on the observation that it is possible to run a polynomial-time classical algorithm, in this case the Goemans-Williamson algorithm \cite{goemans1995improved}, and use partial information from the algorithm's output to inform the QAOA.  
A first idea would be to run the GW algorithm and start the QAOA from a computational basis state corresponding to a cut output by the classical algorithm. 
However, it has been shown that the usual QAOA will not make any progress starting in a  warm computational basis state \cite{cain2023}.
Alternatively, the approach developed in \cite{tate2023bridging,tate2023customMixers} does not start in a computational basis state. They first run the GW algorithm to obtain $n$ unit vectors in $n$-dimensions which have information about the graph. 
These unit vectors are projected down to vectors on a randomly chosen two dimensional plane which are then used to determine the initial state of each qubit in the quantum algorithm.  Furthermore the ``driver'' term in the QAOA is oriented based on the initial qubit states.  
It has been shown that this strategy works to improve the practical performance of the QAOA~\cite{tate2023bridging,tate2023customMixers}.
For a related warm start with some performance guarantees and bounds see~\cite{tate2024guarantees,feeney2024better, egger2021warm}.

In this paper, we make additional modifications which further improve the strategies discussed in previous studies. We explore the use of tree parameters as starting points for optimization. We find that ascending from tree parameters works well in certain cases, although often the improvements from this are too small to be worth the trouble. We also propose modifying the GW warm-start: %{\color{red}we line up} the GW unit vectors corresponding to the qubits' initial states to optimize performance at $p=0$. \sgc{how about ``
we rotate the projected GW  vectors in the 2-dimensional plane to optimize the warm-started QAOA’s performance at $p=0$. Finally, we explore the combination of using tree parameters together with our modified GW warm-start. Unexpectedly, we find that tree parameters continue to work well in this setting.
We explore large graphs, with as many as $512$ vertices, and up to $p=5$ using tensor network techniques. 
These techniques enable us to study the standard QAOA as well as the warm-start setting at these large system sizes.

% \KVK{Should we say that we make slight modifications to these strategies in the intro -- namely ascending from tree parameters on the warm start and choosing alpha? We don't have to go into too much detail, but from Maddie's numerics, we know that this improves performance. Maybe we can signpost to the reader that ``we both combine and make modifications that further improve the strategies discussed in previous studies.''} 
% \textbf{Please  do this.}

\section{Background}

\subsection{A quick review of the standard QAOA}

The QAOA ~\cite{farhi2014qaoa} is designed to find good approximate solutions to the problem of maximizing a classical cost function $C(z)$ defined on bit strings. For example, in a constraint satisfaction problem the
 cost function counts the number of satisfied constraints. The cost function operator $C$ is diagonal in the computational basis,
\begin{equation}
    C \ket{z} = C(z) \ket{z}.
\end{equation}

The standard QAOA applies unitaries that depend on this cost function, with the goal of improving the classical cost function beyond its expected value in the initial state. The initial state $\ket{s}$ takes the form 
\begin{equation}
    \ket{s} = |+\rangle^{\otimes n},
\end{equation} 
where $|+\rangle$ is the $+1$ eigenstate of the single-qubit $X$ operator. Seen in the  computational basis this initial state is the uniform superposition over all bit strings.  Accordingly the expected value of the $C$ in this state is the average of $C(z)$.

The QAOA circuit interleaves a sequence of unitaries depending on the diagonal cost function operator $C$  and the ``mixer'' $B$. The mixer moves between computational basis states, and in the standard QAOA, we take  $B$ as the sum of the single-qubit $X$ operators
\begin{equation}
    B = \sum_i X_i.
\end{equation}
At depth $p$ the QAOA applies a sequence of $2p$ unitaries, which depend on  $2p$ parameters $\boldsymbol{\gamma}=\{\gamma_1,\cdots,\gamma_p\}$ and $\boldsymbol{\beta}=\{\beta_1,\cdots,\beta_p\}$, to the initial state. The resulting, variational state is
\begin{align}
    \ket{\boldsymbol{\gamma}, \boldsymbol{\beta}} = e^{-i\beta_p B} e^{-i\gamma_p C}\dots e^{-i\beta_1 B}e^{-i\gamma_1 C} |+\rangle^{\otimes n}.
\end{align}
% Here $B = \sum_i X_i$ is the sum of the single-qubit $X$ operators, and $C$ is the diagonal cost function operator which we seek to maximize. 
The goal is to find parameters $(\boldsymbol{\gamma}, \boldsymbol{\beta})$ that makes the expectation $\bra{\boldsymbol{\gamma}, \boldsymbol{\beta}} C\ket{\boldsymbol{\gamma}, \boldsymbol{\beta}}$ as large as possible. Note that optimal parameters at $p+1$ can only improve the performance of the optimal parameters at $p$.

\subsection{A quick review of MaxCut and the Goemans-Williamson algorithm}

In this work, we study the QAOA  applied to the \textit{maximum cut} (MaxCut) problem. Consider an undirected graph $G = (V, E)$ with vertex set $V$ of size $n=|V|$ and edge set $E$.  The  MaxCut problem on unweighted graphs is to find a bipartition of the vertices into disjoint sets $S \subseteq V$ and $V \setminus S$, such that number of edges crossing the two sets is  maximized. The edges crossing the two sets are the ``cut'' edges.  For any set $S$ let $Z_i=1$ for $i$ in $S$ and $Z_i=-1$ for $i$ in  $V \setminus S$ then the  cost function 
\begin{align}
\label{eqn:maxcutcostfunction}
C_\mathrm{MC}= \frac{1}{2}
\sum_{\langle i,j\rangle \in E}(1-Z_i Z_j)
\end{align}
counts the number of cut edges associated with the string $Z$. Now the task is to find a string that maximizes this cost function.  The MaxCut problem is NP-complete \cite{Karp1972}, and even approximating it  within a factor of $16/17$ is NP-hard \cite{haastad2001some}. For a given cut, there are two metrics used to quantify performance. First, the cut fraction is defined as the number of cut edges divided by the total number of edges.  Second, the approximation ratio is the number of cut edges divided by the maximum number that can be cut.

In $1995$, Goemans and Williamson gave a method that guarantees an approximation ratio of at least~$0.878$. The Goemans-Williamson (GW) algorithm \cite{goemans1995improved} is a polynomial-time classical approximation scheme for MaxCut based on semi-definite relaxation. The algorithm begins by defining a relaxed cost function where each $Z_i \in \{+1, -1\}$ is replaced by an $n$-dimensional unit vector $\hat{v}_i \in \{ v \in \mathbb{R}^n : \| v\| = 1\}$. Instead of maximizing  $C_{\mathrm{MC}}$ the first task is to maximize 
\begin{align}
\label{eqn:sdp_relaxed_cost}
C_\text{REL}  = \frac{1}{2}\sum_{\langle i,j\rangle \in E}(1-\hat{v}_i \cdot\hat{v}_j).
\end{align}
A solution  $\{\hat{v}^*_i\}_{i\in V}$ maximizing this relaxed objective function $C_\text{REL}$ can be found efficiently in polynomial time using algorithms for semi-definite programming. The solution can be rounded to a good cut: first,  choose a random hyperplane in the $n$-dimensional space, and then for each vertex $i \in [n]$,  set $Z_i=1$ if $\hat{v}^*_i$ is on one side of the hyperplane and $Z_i=-1$ if it is on the other side. Goemans and Williamson proved that this procedure will, for any graph, return a cut with at least a $0.878$ approximation ratio in expectation over hyperplane choices. It was later proven that the approximation ratio of $0.878$ is worst-case optimal under the \textit{unique games conjecture}~\cite{khot2002power}.

\section{Tree parameters work well on random 3-regular graphs \label{sec:tree_parameters}}

In order to run the QAOA, good parameters $(\gambf, \betbf)$ must be  found. This can be done on the fly while running a quantum computer or in advance by running an expensive classical optimization.   Seeking to mitigate the cost of finding good parameters, Brand\~ao et al.~\cite{brandao2018fixed} demonstrated that for fixed parameters, the quantum expectation of the cost function concentrates on typical instances drawn from a distribution. Their work suggested an alternate strategy for running the QAOA. Draw one instance from a given family of graphs, solve for optimal $(\gambf, \betbf)$, and use these parameters on other instances drawn from the same family. This protocol demonstrated promising performance and amortizes the cost of the computational bottleneck (optimizing $(\gambf, \betbf)$) to zero inversely with the number of instances evaluated. It was also shown that parameters, which worked well at one size of graphs, also worked well on larger graphs from the same distribution.

\begin{figure*}[ht]
\centering
\includegraphics[scale = 0.43]{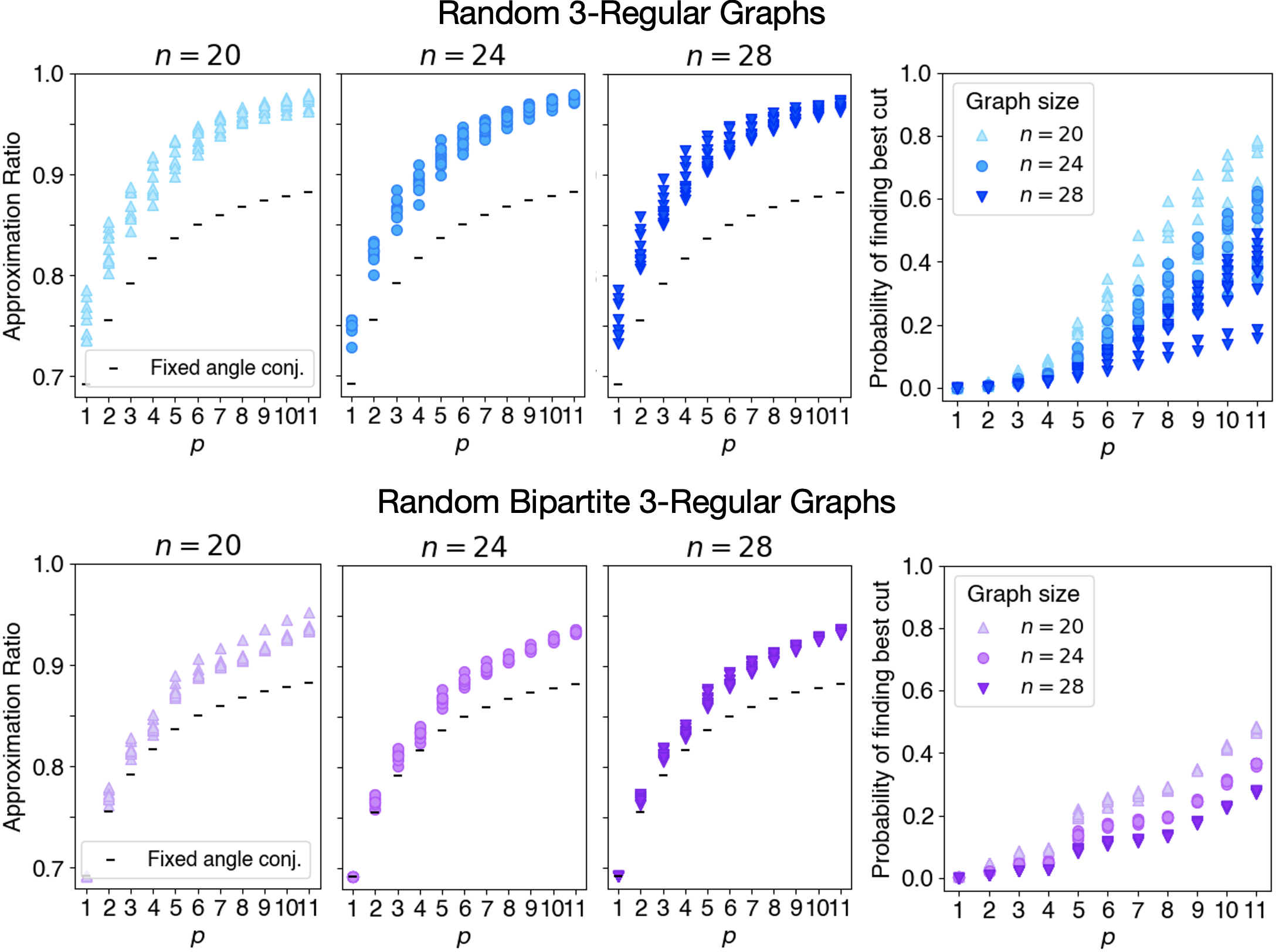}
\caption{\emph{Tree parameter performance on 3-regular graphs at sizes $20$, $24$ and $28$.}  We look at random 3-regular graphs with a single best cut (top)  and random bipartite 3-regular graphs (bottom). 
At each size and graph type, we display $10$ random instances. All of the data lies above the fixed angle conjecture bound $f_{p,3}^{\mathrm{tree}}(\boldsymbol{\gamma_*},\boldsymbol{\beta_*})$ (black dashed lines). The right panel displays the probability of finding the best cut. For both graph types, the probability of finding the best cut decreases with system size. 
}
\label{fig:FigureTreeParams}
\end{figure*}

A line of work has since opened around the observation that there exist easy-to-find and size-independent good parameters that can be used on random instances drawn from a family of graphs \cite{wurtz2021fixed, shaydulin2023parameter, sureshbabu2024parameter}.  Relevant to our work on 3-regular graphs  is the use of parameters computed according to the \textit{Fixed Angle Conjecture} \cite{wurtz2021maxcut}:  consider a $D$-regular 
tree of radius $p$, with central edge $\langle u,v\rangle$.   This is equivalently a $p$-local edge neighborhood of a $D$-regular graph with girth greater than $2p+1$, \emph{i.e.,} a neighborhood containing all vertices within graph distance $p$ of an edge $\langle u,v\rangle$.  Write the QAOA expectation of the central edge of this tree as
\begin{align}
f_{p,D}^{\mathrm{tree}}(\boldsymbol{\gamma},\boldsymbol{\beta}) = \frac{1}{2}\langle \boldsymbol{\gamma},\boldsymbol{\beta}|(I-Z_uZ_v)|\boldsymbol{\gamma},\boldsymbol{\beta}\rangle.
\end{align}
Then the fixed angle conjecture is:

\begin{conjecture*}[\cite{wurtz2021maxcut}]
Let $\mathcal{G}_D$ denote the family of $D$-regular graphs of all sizes. Let $(\boldsymbol{\gamma}_*,\boldsymbol{\beta}_*)$ denote the optimal parameters of the expectation $f_{p,D}^\mathrm{tree}$. Then the depth-$p$ QAOA evaluated at the optimal tree parameters $(\boldsymbol{\gamma}_*,\boldsymbol{\beta}_*)$ will have an approximation ratio for any graph $G\in \mathcal{G}_D$  at least as great as the optimal value of $f_{p,D}^\mathrm{tree}$, i.e.,
\begin{align}
\min_{G\in \mathcal{G}_D} \frac{\langle \boldsymbol{\gamma}_*,\boldsymbol{\beta}_*|C_{\mathrm{MC}}(G)|\boldsymbol{\gamma}_*,\boldsymbol{\beta}_*\rangle}{|\mathrm{MaxCut(G)|}} \ge f_{p,D}^\mathrm{tree}(\boldsymbol{\gamma}_*,\boldsymbol{\beta}_*).
\end{align}
\end{conjecture*}

We emphasize that this conjecture is posited for any graph $G\in \mathcal{G}_D$, regardless of whether or not it is locally tree-like. 
Wurtz and Love \cite{wurtz2021maxcut} prove this conjecture for $D=3$ at $p=1$ and $p=2$. Wurtz and Lykov \cite{wurtz2021fixed} find the optimal parameters for each tree graph up to $p=11$, and they also numerically verify this conjecture up to $p=11$ by considering all 3-regular graphs up to size $16$. At $p=11$ the expected value of the tree graph edge is $0.8828$.  This exceeds the (general purpose) GW $0.878$-guarantee, though there exists a version of GW that is specialized for 3-regular graphs and attains a worst-case approximation ratio of $0.9326$ \cite{halperin2004max}.

Our numerical studies here further support the conjecture. We study graphs at sizes of $20$,  $24$,  and $28$  by simulating the full Hilbert space. 
At each size we toss $10$ random 3-regular graphs, which we post-select to have a single best cut. 
This tightens the distribution and is presumably more difficult than a general random graph. 
We also toss $10$ random bipartite 3-regular graphs.
We evaluate the performance of the QAOA on these graphs using the tree parameters from \cite[Table 1]{wurtz2021fixed}, which extend out to $p$ of $11$.  
See Figure~\ref{fig:FigureTreeParams}. All of our data support the fixed angle conjecture which predicts that  all approximation ratios lie above $f_{p,3}^{\mathrm{tree}}(\boldsymbol{\gamma_*},\boldsymbol{\beta_*})$.

Beyond supporting the fixed angle conjecture, our data shows that the tree parameters give surprisingly high approximation ratios. Consider the data at $p=11$ where the approximation ratios are not far from $1$.  The associated tree graph has $8190$ vertices and in no sense is it a subgraph of a $28$ vertex graph.  It wraps around the $28$ vertex graph many times. It is also remarkable that as a function of $p$ the approximation ratios increase using tree parameters.  The QAOA is a variational algorithm, and for any given graph, optimal parameters' performance can only increase as depth increases. But the tree parameters were not necessarily chosen to be optimal for that specific graph.  

For large $n$, random 3-regular graphs have $O(1)$ triangles, squares, etc. Thus for any fixed $p$, the fraction of edges whose $p$-neighborhoods are trees goes to $1$ as
$n$  goes to infinity. So here the optimal parameters are the tree parameters.
 The maximum cut of a random 3-regular graph for large $n$ is a proper fraction of the number of edges, so for large $n$ the approximation ratios (colored symbols) on the top of Figure~\ref{fig:FigureTreeParams} will lie above the optimal expected cut fraction (dashes). For bipartite graphs, the maximum cut is the number of edges, so for large $n$, the colored symbols in the bottom pane of  Figure~\ref{fig:FigureTreeParams} will lie on the dashes.

One can also gauge how well tree parameters perform, by looking at the probability of finding the best cut after running the QAOA with the tree parameters; see the right panels in Figure~\ref{fig:FigureTreeParams}.  For $p \gtrsim 5$ these probabilities are so high that, with a few runs of the quantum computer, the best cut will be found. The probability to find the best cut decreases with system size, and we believe this probability is decreasing exponentially as the size grows.

\begin{figure*}[hp]
\centering
\includegraphics[scale = 0.36]{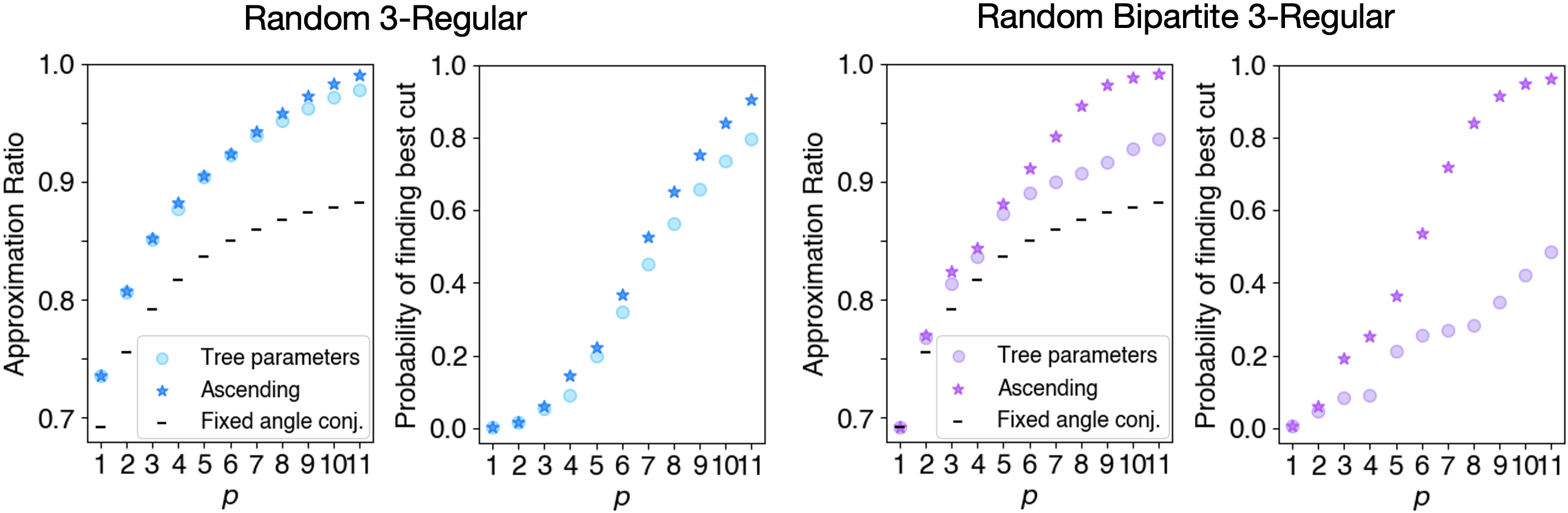}
\caption{
\emph{Ascending from tree parameters.}  
We display one $n=20$ random 3-regular graph (left) and one random bipartite graph (right). We optimize from tree parameters and plot the resulting approximation ratio and probability of finding the best cut.  In both cases we see some improvement, with the bipartite instance showing more. The behavior shown in the two graphs is typical in that other random instances demonstrate similar behavior. 
}

\label{fig:FigureAscending}
\end{figure*}

These numerical studies demonstrate that the tree parameters do a good job on graphs which are not locally tree-like. However, one might also consider using these parameters as the initial guess for a parameter optimization. To test this strategy, we can ask how much improvement is seen after ascending from tree parameters. In Figure~\ref{fig:FigureAscending}  we show  the  results for both  random $3$-regular graphs and random bipartite $3$-regular graphs at size $n=20$. While both graph types exhibit some improvement when optimizing from tree parameters, the bipartite case has more improvement and suggests that sometimes ascending from tree parameters is a good strategy.

Later, in Section~\ref{sec:NumericalResults}, we use tensor network methods to analyze graphs of size up to $n = 512$ and depths up to $p=5$.  We show the approximation ratios at tree parameters in Figure \ref{fig:TensorNetworkNumerics} and find that these are about the same as their values at much lower system sizes.

\section{Warm-starting QAOA with help from Goemans-Williamson}\label{subsec:warmstartQAOA}

The practice of ``warm-starting'' an algorithm involves giving it an approximate solution to a problem, with the goal of increasing its performance by providing a strategic starting point from which it may improve the solution further. When warm starting the QAOA  different choices of initial state  can drastically impact  subsequent performance. It was shown in~\cite{cain2023} that initial states consisting of a single classical bit-string will generally stall the algorithm, providing no further improvements. Non-classical products state inputs have been shown  to perform better, especially when paired with appropriate modifications to the QAOA mixing operator $B$. Hereafter, we will directly follow the warm start approach of \cite{tate2023customMixers}, which we  refer to as the ``GW warm-start", to leverage information from the Goemans-Williamson SDP relaxation for MaxCut  and modify the QAOA.

A solution to the relaxed GW problem of equation~\ref{eqn:sdp_relaxed_cost} consists of normalized vectors $\hat{v}^*_i \in \mathbb{R}^n$, one for each vertex $i \in [n]$. Project these solution vectors onto a random 2-dimensional plane, and pick a random basis for the plane which then defines the $x$-direction and the $z$-direction.   The projected vector coming from  $\hat{v}^*_i$ has an angle $\theta_i$ with the $z$-direction.  Normalize each projected vector and place each  normalized vector on the Bloch circle.    
For each vertex $i \in [n]$ there is a corresponding qubit state
\begin{align}
|\theta_i\rangle = \cos\left( \theta_i/2\right)|0\rangle + \sin\left( \theta_i/2\right)|1\rangle.
\end{align}
The initial state of the GW warm-start QAOA $\ket{\boldsymbol{\theta}}$ is the tensor product of the $\ket{\theta_i}$ single-qubit states for each vertex $i \in [n]$, 
\begin{equation}
    |\boldsymbol{\theta}\rangle = |\theta_1\rangle\otimes\cdots\otimes|\theta_n\rangle.
\end{equation}
Also modify the mixing operator 
\begin{align}
    B = \sum_i B_i \label{eq:mixer}
\end{align}
such that each $B_i$ points in the direction of its corresponding vector on the Bloch circle,
\begin{equation}
    B_i = \sin(\theta_i) X_i + \cos(\theta_i) Z_i.
\end{equation}
Just as with the standard QAOA, the goal of the GW warm-start QAOA is to maximize the expectation value of the cost function $C_\mathrm{MC}$ in the variational state $\ket{\boldsymbol{\gamma}, \boldsymbol{\beta}}$. The crucial difference between the two is that the GW warm-start's initial state is $|\boldsymbol{\theta}\rangle$ and mixer is $B$ coming from equation~\eqref{eq:mixer}.

Note that each $\ket{\theta_i}$ is a  $+1$ eigenstate of the corresponding $B_i$.
This choice of $B_i$ ensures that the adiabatic guarantee is retained with the input state $|\boldsymbol{\theta}\rangle$, as is the case with the standard QAOA. This means that for any fixed instance there are parameters as $p \to \infty$ that guarantee that the quantum algorithm will go to the maximum of the cost function.  However, we always work with finite $p$; while the adiabatic guarantee holds, its relevance is not clear in the finite $p$ regime. 

In Appendix~\ref{app:p0_performance_guarantee} we lower-bound the $p=0$ performance of this warm-start QAOA, as in~\cite{tate2023customMixers}. At $p=0$ we measure the state $|\boldsymbol{\theta}\rangle$ in the computational basis and use the observed string to give the cut. For the standard QAOA, if we measure the initial state at $p=0$, we get a random string, and the expected cut fraction is $1/2$. The warm-start initial state is informed by the GW SDP solution. 
For a graph $G$, let $\mathscr{C}_\text{GW}$ be the expected GW cut value, where the expectation is over all GW hyperplane roundings. We provide a self-contained proof that the $p=0$ the warm-start QAOA achieves a cut value of at least $\frac{3}{4}  \mathscr{C}_\text{GW}$~\cite{tate2023customMixers}. 

\subsection{A modification at $p=0$}
\label{sec:alpha}
The $\frac{3}{4}\mathscr{C}_\text{GW}$ guarantee discussed above assumes that the $x$- and $z$-axes of the Bloch circle are randomly oriented.   We can improve the $p=0$ performance by optimizing this orientation.  We find this simple classical optimization at $p=0$ also benefits the $p>0$ performance.
For a single qubit with angle $\theta_i$ a measurement of $Z_i$ has a quantum expectation of $\cos \theta_i$. This means that the quantum expectation of the cut value in  the initial product state is 
\begin{equation}
 \sum_{\langle i , j \rangle \in E}   \frac{1}{2}  \Big( 1 - \cos(\theta_i) \cos(\theta_j) \Big).
\end{equation}
Rotating the axes by $\alpha$ is equivalent to shifting each $\theta_i$ by $\alpha$, and this expected cut value becomes
\begin{equation}
 \sum_{\langle i , j \rangle \in E}   \frac{1}{2}  \Big( 1 - \cos(\theta_i+\alpha) \cos(\theta_j+\alpha) \Big).
\end{equation}
This function of $\alpha$ can be classically optimized before going to the quantum computer, improving the $p=0$ performance over that of a random orientation.  
Therefore, part of our strategy is to do this optimization and keep the newly oriented angles $\theta_i$ when we go to higher $p$.  
We motivate this strategy in the next section by considering the special case of bipartite graphs, for which the $p=0$ optimization already produces the optimal cut.

\subsection{Bipartite graphs and motivation for optimizing at $p=0$} 

 Bipartite graphs can be used to illustrate the limitations of the standard QAOA and the benefits of the GW warm start. Consider a large random 3-regular graph and a large random \textit{bipartite} 3-regular graph, and run the standard QAOA on each at low depth $p$.  The QAOA's performance is determined by the local neighborhoods of the edges. For both graphs here, these neighborhoods are trees with high probability, so the cut value produced by the QAOA is the same for both graphs. In the bipartite case, the optimal cut consists of all the edges, whereas the optimal cut value in the fully random  case is only a fraction of the number of edges.  We conclude that the approximation ratio of the QAOA at low depth on random bipartite graphs is upper bounded by the MaxCut fraction of the fully random case.
 
This kind of locality argument limiting the QAOA performance does not apply to the GW algorithm, which starts by optimizing a non-local relaxed cost function. Locality arguments, such as those provided in \cite{farhi2020quantumTypical, farhi2020quantumWorst}, therefore cannot be used to bound performance of the GW warm-start QAOA.  In fact, the GW algorithm exhibits perfect performance on bipartite graphs, that is, $\mathscr{C}_\text{GW} = |E|$. To see this, consider the $n$-dimensional solution $\{\hat{v}^*_i\}$ that maximizes the GW relaxed cost function of equation~\eqref{eqn:sdp_relaxed_cost}.  
The best solution is when all vectors  corresponding to vertices $i$ in one independent set point in some direction $\hat{v}^*_i = \hat{u}$, while the vectors corresponding to $j$ in the other set point in the opposite direction $\hat{v}^*_j = -\hat{u}$.
Under any choice of hyperplane, the rounding procedure finds the maximum cut, which bipartitions the vertices into the two independent sets.

In the GW warm-start, these SDP solution vectors  $\{\hat{v}^*_i\}$ are projected onto a random 2-dimensional plane and normalized. Under any choice of 2-dimensional plane, the projected vectors point in opposite directions, $\hat{r}$ and $-\hat{r}$, on the Bloch circle. Define $\theta$ as the angle between $0$ and $\pi$ that $\hat{r}$ or $-\hat{r}$ makes with  the positive  $z$-axis. 
When $\theta$ is $0$ the initial state $|\boldsymbol{\theta}\rangle$ is a computational basis state corresponding to the best cut, meaning at $p=0$ we inherit the perfect performance of the GW algorithm. 
When $\theta$ is $\frac{\pi}{2}$  the warm-start QAOA's initial state is a product of single qubit states pointing in the $+x$ or $-x$ directions. This is a uniform superposition of all computational basis states with varying signs, and the expected cut fraction is $\frac{1}{2}$.
For a general  $\theta$  we have  that $\bra{\boldsymbol{\theta}} C_\text{MC} \ket{\boldsymbol{\theta}}$ is $\frac{1}{2} + \frac{1}{2} \cos^2\theta$, and the average of this over $\theta$ is  $\frac{3}{4}$. Therefore, the bipartite case is the worst case for the $p=0$ bound  of $\frac{3}{4}  \mathscr{C}_\text{GW}$, which we previously stated; see appendix A1.

In the bipartite case, it is especially useful to optimize the orientation of the $2$-dimensional unit vectors on the Bloch circle.  
This means we want $\theta=0$, i.e., $\hat{r}$ points along the $z$ axis, and we pick $\alpha$ of 
the previous subsection such that $\alpha + \theta_i = 0$.  
In this case, the initial state is a computational basis state that gives the perfect cut--there is no need to go to higher $p$.  We note that previously, we warned the reader that warm starting from a computational basis state is a poor strategy in general because the QAOA gets stuck \cite{cain2023}. Here that point is moot, as there is no room for improvement from the best cut.

\subsection{Optimization at $p=0$ beyond bipartite graphs}

\begin{figure*}[htbp]
    \centering
    \begin{subfigure}[t]{0.48\textwidth}
        \centering
        \begin{tikzpicture}
            \node[inner sep=0pt] (image) {\includegraphics[width=\textwidth]{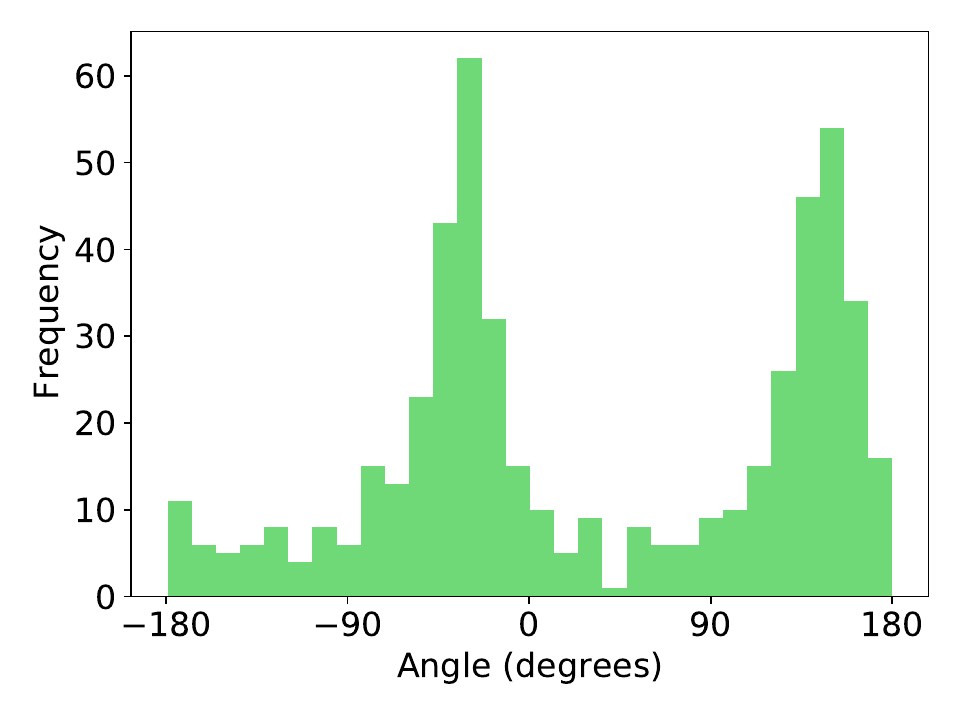}};
            \node[anchor=north west, inner sep=0pt, font=\bfseries] at ([xshift=-5pt,yshift=5pt]image.north west) {
            %(a)
            };
        \end{tikzpicture}
    \end{subfigure}
    \hfill
    \begin{subfigure}[t]{0.48\textwidth}
        \centering
        \begin{tikzpicture}
            \node[inner sep=0pt] (image) {\includegraphics[width=\textwidth]{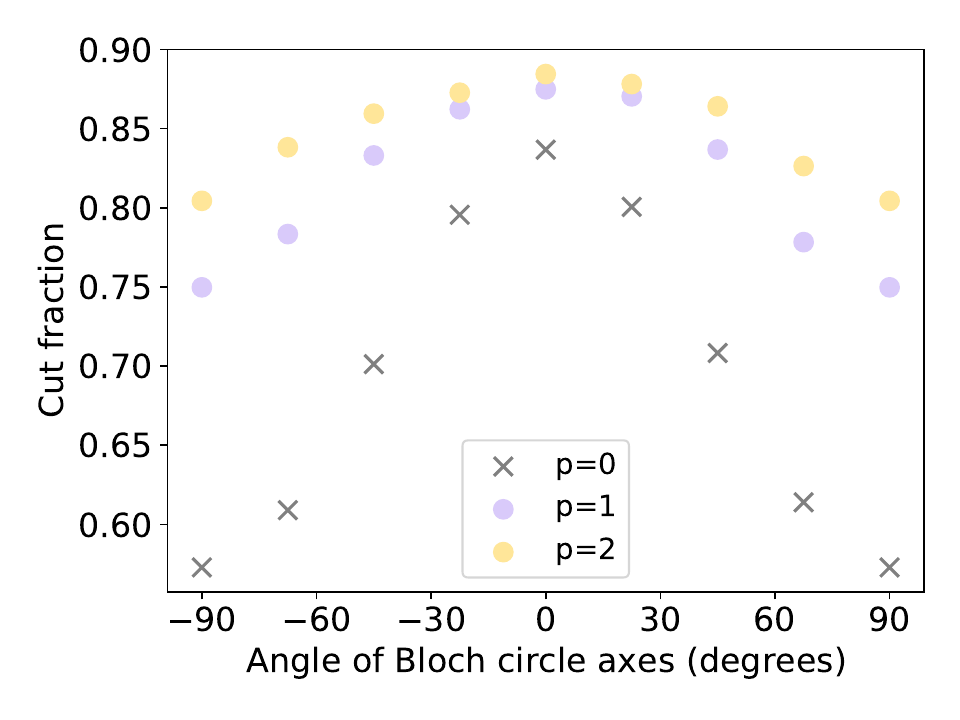}};
            \node[anchor=north west, inner sep=0pt, font=\bfseries] at ([xshift=-5pt,yshift=5pt]image.north west) {
            };
        \end{tikzpicture}
    \end{subfigure}
    \caption{\textit{Dependence on Bloch circle axes.}  Left: for a single random 3-regular graph with 512 vertices, we plot the distribution of angles $\{\theta_i\}$ specifying a warm start, obtained from the Goemans-Williamson SDP solution. Right: for a single random 3-regular graph with 256 vertices, we plot the QAOA performance at $p=0,1,2$, for different choices of the angle of the axes of the Bloch circle.  The angle labeled zero on the figure's $x$-axis is defined to maximize the $p=0$ performance.}
    \label{fig:p=0}
\end{figure*}

We want to carry the above strategy beyond the bipartite case, where the best cut does not include all of the edges.  It is helpful to visualize the distribution of angles $\{\theta_i\}$ specifying the warm start. 
Recall these are obtained from the vectors $\{\hat{v}^*_i\}$ of the SDP solution by projecting onto a random 2-dimensional plane and measuring the angle with respect to a chosen $z$ axis.  In the bipartite case, these vectors were all parallel or anti-parallel, so the angles $\{\theta_i\}$ take exactly two values, separated by 180 degrees.  In Figure \ref{fig:p=0}(a) we illustrate the distribution of angles $\{\theta_i\}$ for the case of a random 3-regular graph with 512 vertices.  Evidently they form two loose clusters separated by approximately 180 degrees.  (The choice of angle zero here is arbitrary.)  Optimizing the orientation of the $z$ axis at $p=0$, i.e., choosing $\alpha$ of Section~\ref{sec:alpha},  will put these clusters roughly along the $z$ axis.

 It is natural to wonder whether the optimal orientation of the axes at $p=0$ also benefits the performance of the QAOA at $p>0$.  The answer for low $p$ is illustrated in Figure \ref{fig:p=0}(b), with data for a single random 3-regular graph with 256 vertices.  The gray crosses indicate the expected cut fraction yielded by the QAOA at $p=0$, obtained by measuring the qubits in the computational basis, without running a circuit.  The performance is plotted as a function of the angle of the axes of the Bloch circle. We define $0$  to be the value that maximizes the $p=0$ cut fraction.  (The gray crosses therefore peak at angle 0 by definition.)   We see there is large variation in cut fraction with respect to the choice of angle of the axes.  We also plot performance of the QAOA at $p=1, 2$.  We then see the variation with respect to the choice of axes is diminished but still significant.  

 This data supports our proposal to modify the warm start by optimizing the choice of axes for the Bloch circle at $p=0$, using this same warm-start for higher $p$ as well. Note that a similar optimization for choosing the GW cut is also possible and seems to improve the GW performance slightly. 
 
\section{Numerical Results on Large Graphs}\label{sec:NumericalResults}

\begin{figure*}[!ht]
\centering
\includegraphics[scale = 1]{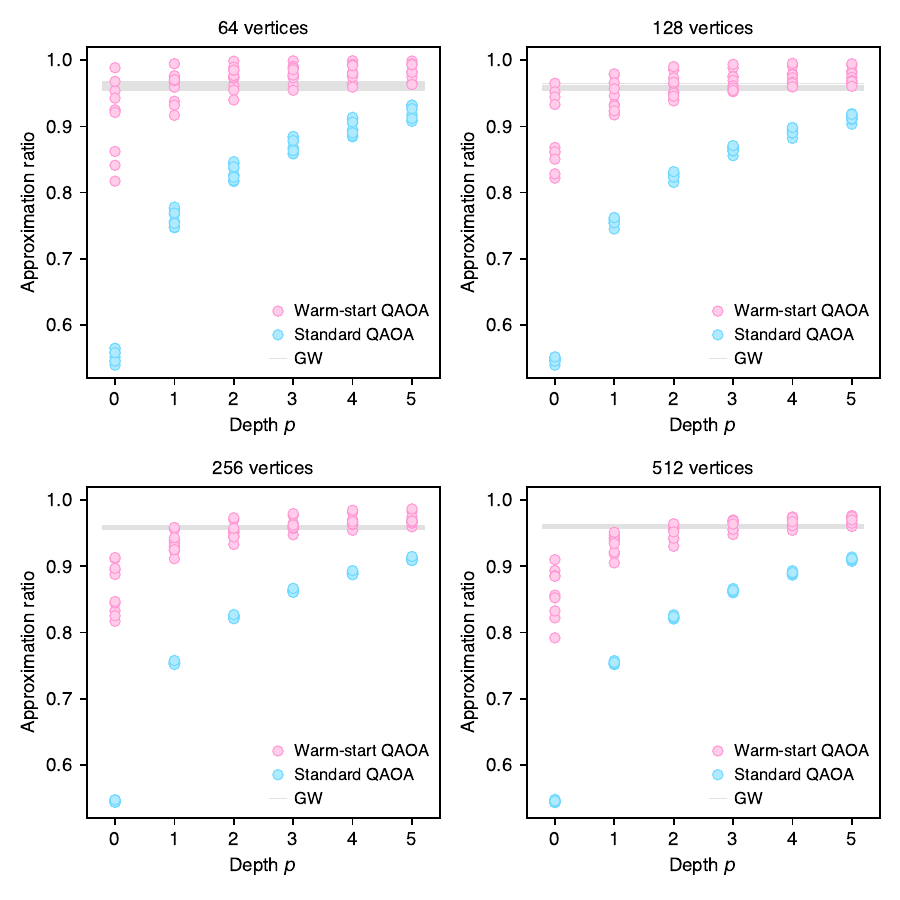}
\caption{\emph{Standard and GW warm-start QAOA on random 3-regular graphs of 64, 128, 256 and 512 vertices.} 
We generate ten graphs at each system size and show performance for each graph for different values of the depth $p$. 
For the standard QAOA ({\color[rgb]{0.4392, 0.8509, 1}blue}), we plot the approximation ratios achieved at tree parameters. Further optimization does not significantly improve performance.
For the GW warm-start ({\color[rgb]{1, .6, 0.8392}pink}), we optimize the  orientation $\alpha$ at $p=0$, and then use the tree parameters.  
The $p=0$ performance (when the QAOA unitary is trivial) is shown for reference.
The thick grey line shows the tiny spread of the expected GW cuts for the ten instances at each size. }
\label{fig:TensorNetworkNumerics}
\end{figure*}

In this section, we present data  for  both  the standard  and GW warm-start QAOA at system sizes larger than those which can be analyzed by direct  simulation in the full Hilbert space, as in Section~\ref{sec:tree_parameters}.  For the data in this section we are  able to go to $p=5$ before the numerical techniques become too costly. We leverage the locality of the QAOA to compute the cost function on each edge independently (see Appendix~\ref{sec:neighborhoods}).
We then employ tensor network methods which have been explored in the literature. 
We use the following package to generate the circuits~\cite{Luo2020yaojlextensible}, then recast the circuit into a tensor network~\cite{Kalachev2021, Pan2022, Markov2008}. 
We optimize the contraction order of the tensor network~\cite{10.21468/SciPostPhys.7.5.060, Gray2021, Kalachev2021, Pan2021} using an open-source package~\cite{Liu2022}.

The GW warm-start QAOA has the same locality structure as the standard QAOA. 
The initial state is a product state, the mixing operator $B$ is still a sum of one qubit operators, and the cost function operator is unchanged.  This allows us to use the same tensor network techniques for this seemingly more complicated algorithm. 

In Figure~\ref{fig:TensorNetworkNumerics} we show data obtained by sampling ten random $3$-regular graphs at sizes $n=64, 128, 256$ and  $512$. We do not post-select on there being a single best cut as we did in the earlier data.  For the standard QAOA (in {\color[rgb]{0.4392, 0.8509, 1}blue}), we use the tree parameters of Wurtz and Lykov~\cite{wurtz2021fixed}. For $p\le 4$ and graph sizes up to $n=256$, our numerical results match those obtained by Wurtz and Lykov. Optimizing from the tree parameters produces tiny improvements, so we show only the tree parameter data. The ability to choose parameters in advance without further optimization is advantageous especially when running on hardware.  Note that the values of the approximation ratios are very similar to those at sizes $20, 24$ and $28$ for the values of $p$  that are shown.

We also show the GW warm-start performance for the same instances.  
At $p=0$ we do the optimization of $\alpha$ described in Section~\ref{subsec:warmstartQAOA}.  
This makes the  approximation ratios start high, which is why we dub this a ``warm start.''  
For $p>0$ we use the same tree parameters $(\mathbf{\gamma}_*,\mathbf{\beta}_*)$ as were used for the ordinary QAOA. The results are displayed in Figure~\ref{fig:TensorNetworkNumerics} in {\color[rgb]{1, .6, 0.8392}pink}.  We also considered further optimization starting from the tree parameters, and we find that this gives an approximation ratio improvement of roughly 0.01 at $p=4$ for all sizes considered, which we do not display. It is notable that the tree parameters, which were chosen for the ordinary QAOA, continue to perform so well for the warm-start QAOA.
We plot the GW expected cut value for each instance, which results in  the slightly thickened grey line.  While the warm-start QAOA does pass this line, we do not claim that this demonstrates quantum advantage.

\section{Summary}

There are many heuristic classical algorithms for combinatorial search problems which work well in practice.  It is unknown if quantum computers can offer any advantage over these classical algorithms.  Proving bounds on the performance of quantum optimization algorithms is notoriously difficult.  However, we can still explore strategies for running quantum algorithms in regimes where quantum hardware might reveal insights that cannot be determined classically.

In this paper we focused on the QAOA, an all purpose optimizer whose full power is unknown.  We restrict our attention to MaxCut on 3-regular graphs and on problem sizes of up to $512$ vertices.
We offer the following strategies for running on hardware that is fault tolerant and capable of handling instances with hundreds of vertices or more.  As we explain in the paper, these strategies are not original to us, although we have attempted to improve and combine aspects of them.\\

\noindent
\textbf{Task}: Given a $3$-regular graph $G$ with $n$ vertices, find a cut with value close to the MaxCut.\\

\noindent
\textbf{Given}:  A quantum computer with high-fidelity gates capable of running the QAOA to depth $p$ on a problem instance of size $n$.\\

\noindent
\textbf{Procedure}: Consider a $3$-regular tree with radius $p$. Classically determine the optimal parameters $(\boldsymbol{\gamma}_*, \boldsymbol{\beta}_*)$ for the cost function on the central edge. (Or look them up.) This is a one-shot calculation, and the parameters $(\boldsymbol{\gamma}_*, \boldsymbol{\beta}_*)$ will be used for all instances of any size at this value of $p$. At this point, one can choose between a few strategies:

\begin{itemize}
\item[]  \textbf{Strategy 1}: 
\begin{itemize}
    \item Run the level-$p$ QAOA on $G$ with parameters $(\boldsymbol{\gamma}_*, \boldsymbol{\beta}_*)$.

    \item Hardware resources permitting, use the parameters $(\boldsymbol{\gamma}_*, \boldsymbol{\beta}_*)$ as the starting point of a parameter search to look for a better cut on $G$.  This may require many calls to the quantum hardware but may improve performance.
\end{itemize}

\item[] \textbf{Strategy 2}:  

\begin{itemize}
    \item Find $n$ unit vectors that optimize the relaxed cost function of the Goemans-Williamson algorithm on $G$.  

    \item Sample a random two-dimensional plane and project these vectors onto this plane.  Normalize each projected vector and place the result on the Bloch circle. These points can be associated with a quantum state given as a product of single-qubit states defined by the corresponding points on the Bloch circle. 

    \item Perform a classical optimization by rotating the Bloch circle so that the expectation of the MaxCut cost function in the quantum state is maximized. The resulting state will serve as the initial state of the QAOA. Modify the driving operators $e^{-i\beta X_j}$ so that they align with the qubits of the initial state, \emph{i.e.}, $X_j \rightarrow Z_j \cos \theta_j + X_j \sin \theta _j$, where $\theta_j$ is the angle of the $j$th vector with the $+Z$-axis.  

    \item Run this modified QAOA to depth $p$ using the parameters $(\boldsymbol{\gamma}_*, \boldsymbol{\beta}_*)$.

    \item Hardware resources permitting,  use the parameters $(\boldsymbol{\gamma}_*, \boldsymbol{\beta}_*)$ as the starting point of a parameter search to look for a better cut on $G$. This may require many calls to the quantum hardware but may improve performance.
\end{itemize}
\end{itemize}

We imagine using these strategies at higher $p$ than what we have explored numerically. Looking at Figure \ref{fig:TensorNetworkNumerics}, it is not clear whether the standard QAOA with tree parameters will eventually surpass the GW warm-start QAOA at larger values of $p$. 
Both strategies show promise in various regimes. A key area of future investigation is to apply these strategies to other combinatorial optimization problems, where the analogue of tree parameters is not immediately clear. %and benchmark their performance against state-of-the-art classical solvers.

\section*{Acknowledgements}
M.C. acknowledges support from DOE CSG award fellowship (DESC0020347). D.R. acknowledges support from NTT (Grant AGMT DTD 9/24/20). E.T. acknowledges funding received from DARPA 134371-5113608, and DOD grant award KK2014. Sw.G. acknowledges support from DARPA  HR001120C0046. K.V.K. acknowledges support from the Fannie and John Hertz Foundation and the National Defense Science and Engineering Graduate (NDSEG) fellowship. 

We thank Johnnie Gray, Robbie King,  Jinguo Liu, Reuben Tate, Ben Villalonga, and Jonathan Wurtz for helpful discussions.

\appendix
\section{Theoretical guarantee for warm-started QAOA at p=0}\label{app:p0_performance_guarantee}

Here we evaluate the $p=0$ performance of the warm-start QAOA.  By $p=0$ we mean that we measure the state $|\boldsymbol{\theta}\rangle$ in the computational basis and use the observed string to get the cut value. This quantum expectation needs to be averaged over the choice  of 2-dimensional plane and the orientation of the $x$ and $z$ axes.

A solution to the GW SDP relaxation consists of normalized vectors $\hat{v}^*_i \in \mathbb{R}^n$, one for each vertex $i$. In the GW algorithm these are rounded to give a cut: select a random line in $\mathbb{R}^n$ and project each vector $\hat{v}^*_i$ onto this line. Upon normalization, each vector is either $+1$ or $-1$, and this partition defines the GW cut. The expected value over the choice of the line, $\mathscr{C}_\text{GW}$, is the sum of the probabilities of each edge being cut during the rounding: 
\begin{equation}
    \mathscr{C}_\text{GW} = \sum_{\langle i , j \rangle \in E} \frac{1}{\pi} \cos^{-1} \left( \hat{v}^*_i \cdot \hat{v}^*_j \right).
    % = \sum_{\langle i , j \rangle \in E} \frac{1}{\pi} \Embb \left[ \cos^{-1} \left( x_i \cdot x_j \right) \right],
\end{equation}

For the GW warm-start QAOA, we project the SDP solution vectors $\hat{v}^*_i$  onto a random 2-dimensional plane. 
After normalization the projected vectors $\hat{x}_i \in \mathbb{R}^2$ define the warm-start QAOA's initial state. 
If we were to project each 2-dimensional vector $\hat{x}_i$ onto a random line in $\mathbb{R}^2$, this would be is equivalent to performing the full hyperplane rounding of the GW algorithm. Therefore, the GW cut value can also be expressed in terms of the projected vectors $\hat{x}_i$ 
\begin{equation}
\label{eqn:cutGWintermsofx}
    \mathscr{C}_\text{GW}  
    = \sum_{\langle i , j \rangle \in E} \frac{1}{\pi} \Embb \left[ \cos^{-1} \left( \hat{x}_i \cdot \hat{x}_j \right) \right],
\end{equation}
where the expectation value is taken over all 2-dimensional planes.
 For fixed $\hat{x}_i$, the warm-start's performance at $p=0$ is the quantum expectation value of $C_\text{MC}$, defined in equation~\eqref{eqn:maxcutcostfunction}, which is
\begin{equation}
    \bra{\boldsymbol{\theta}} C_\text{MC} \ket{\boldsymbol{\theta}} 
    = 
    \sum_{\langle i , j \rangle \in E}   \frac{1}{2}  \Big( 1 - \cos(\theta_i) \cos(\theta_j) \Big).
\end{equation}

In this expression, we can re-express $\theta_j$ as $\theta_j = \theta_i + \theta_{ij}$, where $\theta_{ij}$ is defined as the difference between the two angles. For each choice of a random 2-dimensional plane, averaging over the choice of the $z$-direction is equivalent to averaging over $\theta_{i}$. Taking the expectation over the choice of the $z$-direction gives
\begin{equation}
    \Embb \left[ \bra{\boldsymbol{\theta}} C_\text{MC} \ket{\boldsymbol{\theta}} \right] 
    =  \Embb \Big[ \sum_{\langle i , j \rangle \in E}   \frac{1}{2}  \Big( 1 - \frac{1}{2} \cos\left(\theta_{ij} \right) \Big) \Big] .
\end{equation}
where $\Embb$ is over the choice of 2-dimensional planes.
We can now relate the expected value of the cost function at $p=0$ to the expected GW cut value $\mathscr{C}_\text{GW}$. First we re-express the right side of equation~\eqref{eqn:cutGWintermsofx} in terms of $\theta_{ij}$,
\begin{equation}
    \mathscr{C}_\text{GW}  
    = \sum_{\langle i , j \rangle \in E} \frac{1}{\pi} \Embb \left[ \theta_{ij} \right].
\end{equation}
We relate these two expectation values using the following inequality:
\begin{equation}
    \Embb \left[  f(x) \right]  \geq \left[ \min_t \frac{f(t)}{g(t)} \right] \Embb \left[ g(x)\right] ,
\end{equation}
where the functions $f$ and $g$ are both assumed to be non-negative. 
The expected value, over choice of two dimensional planes and axis orientation, of the quantum expectation value at $p=0$ can now be lower bounded
\begin{equation}
    \Embb \left[  \frac{1}{2}\Big(1-\frac{1}{2}\cos(\theta_{ij})\Big) \right]
    \geq 
    \min_t \frac{\frac{1}{2}\big(1-\frac{1}{2}\cos(t)\big)}{t/\pi} \hspace{2mm} \Embb \left[ \frac{\theta_{ij}}{\pi}\right].
\end{equation}
The minimum over $t$ is achieved at $t=\pi$ and equals $3/4$. Thus

\begin{equation}
    \Embb \left[ \bra{\boldsymbol{\theta}} C_\text{MC} \ket{\boldsymbol{\theta}} 
    \right]  \geq  \frac{3}{4} \mathscr{C}_\text{GW}.
    % \langle C_\text{MC} \rangle_{p=0} \geq 
    % \frac{3}{4} \sum_{\langle i , j \rangle \in E} \frac{1}{\pi} \Embb (\theta_{ij}),
\end{equation}

\section{Quantum expectation as a sum over neighborhoods}\label{sec:neighborhoods}

For both the standard QAOA and warm-start QAOA,  the expected cost function is a sum over contributions from subgraphs, each of which is the light cone of an edge.  This can be used to great advantage for numerics since expectation values can be computed as a sum of computationally inexpensive local terms. 

Consider the variational state $\ket{\boldsymbol{\gamma}, \boldsymbol{\beta}} = U\ket{\phi}$. Here, $U = U(\boldsymbol{\gamma}, \boldsymbol{\beta})$ is either the standard or warm-start QAOA unitary, and the initial state $\ket{\boldsymbol{\phi}}$ is a product state with each qubit $i$ pointing in some direction $\ket{\phi_i}$ on the Bloch circle. In  the warm-start QAOA case, this state has $\ket{\phi_i} = \ket{\theta_i}$,  whereas in the standard QAOA case each qubit points along the $x$-direction: $\ket{\phi_i} = \ket{+}$. 

The expected cost function  $\bra{\boldsymbol{\phi}}U^{\dagger} C_{\text{MC}} U\ket{\boldsymbol{\phi}}$  is a sum over edges 
\begin{equation}
    \bra{\boldsymbol{\phi}}U^{\dagger} C_{\text{MC}}  U\ket{\boldsymbol{\phi}} = 
    \frac{1}{2} |E| -  \frac{1}{2} \sum_{\langle i,j\rangle \in E}\bra{\boldsymbol{\phi}}U^{\dagger} Z_i Z_j  U\ket{\boldsymbol{\phi}} .
\end{equation}
For each edge $\langle i,j\rangle$, the term $U^{\dagger}Z_iZ_jU$  is supported on the ``edge neighborhood'' of $\langle i,j\rangle$ because the evolution is local in both the standard and warm-start QAOAs. This edge neighborhood contains all vertices within graph distance $p$ of $\langle i,j\rangle$.  Therefore, when evaluating the term $U^{\dagger}Z_iZ_jU$ in the expected cost function, we only need to consider the vertices in this neighborhood. We can replace $\ket{\boldsymbol{\phi}}$ with only its single qubit components that live in this neighborhood, $|\boldsymbol{\phi}_{\langle i,j\rangle}\rangle$.
With this substitution, the expected cost function can be written as a sum over edge neighborhoods,
\begin{equation}
    \label{eq:edge_sum}
 \bra{\boldsymbol{\phi}}U^{\dagger} C_{\text{MC}}  U\ket{\boldsymbol{\phi}} = 
    \frac{1}{2} |E| -  \frac{1}{2} \sum_{\langle i,j\rangle \in E} \langle \boldsymbol{\phi}_{\langle i,j\rangle} | \hspace{1mm}U^{\dagger} Z_i Z_j  U \hspace{1mm} | \boldsymbol{\phi}_{\langle i,j\rangle} \rangle .
\end{equation}
When evaluating the QAOA's performance at depth $p$, each term can be computed using only the Hilbert space of qubits in the corresponding edge neighborhood. The size of each edge neighborhood grows with $p$, which limits the depth to which we can analyze performance in practice.

\bibliographystyle{myalpha}
\bibliography{sdpbib}

\end{document}